\documentstyle[preprint,aps,prb,epsfig]{revtex}

\draft
\begin{document}

\title{A dimerized spin fluid in a one-dimensional electron system}

\author{Y. Z. Zhang$^{1,2}$, C. Q. Wu$^1$ , and H. Q. Lin$^3$ }

\address{
$^1$Research Center for Theoretical Physics, Fudan University,
Shanghai 200433, China\\
$^2$Surface Physics Laboratory, Fudan University, Shanghai 200433,China\\
$^3$Department of Physics, The Chinese University of Hong Kong,
Hong Kong, China }

\date{\today}
\maketitle

\begin{abstract}
The ground state of a one-dimensional Hubbard model with a bond-charge
attraction $W$ term at half-filling is investigated by the density matrix
renormalization group method. It is confirmed that the spin gap will
be closed at $U>8W$. But the long-range bond order wave
survives even when the spin gap is closed. It indicates
that the ground state is a novel dimerized spin fluid at $U>8W$. By a
charge-spin transformation, it is shown that there should be a dimerized
metallic phase at $U<-8W$. Furthermore, it is found that the Hubbard
interaction $U$ enhances initially the dimerization for a weak bond charge
attraction $W$ whereas it reduces monotonously the dimerization for 
a stronger bond charge attraction $W$.
\end{abstract}

\pacs{PACS:71.30.+h; 71.10.Fd; 71.10.Pm}

The response of correlated electron systems to lattice distortions has been
extensively studied motivated not only by theoretical interest but also by
the discovery of quasi-one-dimensional conductors and 
high-T$_c$ superconductivity.
It is well known that Peierls instability\cite{1.Peierls} could induce 
a lattice dimerization in a one-dimensional(1d) electron-phonon (e-p)
interacting system\cite{2.Heeger} at half filling.
A dimerized system has a long-range bond order wave (BOW).
The BOW could be enhanced by an electron-electron (e-e) interaction,
although a strong Coulomb e-e interaction will always reduce the BOW,
which exists until $U\rightarrow \infty $ \cite{3.Baeriswyl,4.Wu}.
In the limit of very strong on-site
Coulomb interaction, the Peierls instability is usually referred to as the
spin-Peierls instability. Within a spin-Peierls model, the quantum phonons
have been taken into account recently \cite{5.Sandvik,6.Bursill,7.Uhrig} and
a quantum phase transition from a gapless spin-fluid state to a
gapped dimerized phase is found to occur at a nonzero value of the
spin(electron)-phonon coupling. It is believed that the BOW long-range
order will disappear as the spin gap is closed \cite{8.Campbell}. On the
other hand, the BOW has been shown to exist in an interacting electron
model recently. \cite{9.Nakamura,10.Sengupta,11.Egami,12.Fabrizio,13.Qin}
In this paper, we investigate the 1d Hubbard
model with a bond-charge attraction term at half-filling by the density
matrix renormalization group (DMRG) method \cite{14.White} and 
find an interesting phenomenon that 
the BOW long-range order survives even the spin gap is closed
for $U>8W$, which shows that the ground state is a novel dimerized spin
fluid phase. The result is in contradiction to that obtained by a
weak-coupling continuum-limit approach \cite{15.Japaridze}.

The Hamiltonian of the 1d interacting electron system at half-filling
\cite{16.Assaad,17.Daul} we consider is as follows
\begin{equation}
H=H_U+H_W ~,
\end{equation}
where $H_U$ is the usual Hubbard model
\begin{equation}
H_U = -t \sum_{l,\sigma} \left( c_{l,\sigma }^{\dagger}c_{l+1,\sigma }
+ c_{l+1,\sigma }^{\dagger} c_{l,\sigma } \right ) 
+ U \sum_l \left( n_{l\uparrow}-\frac 12\right)
\left( n_{l\downarrow }-\frac 12\right) ~,
\end{equation}
and $H_W$ is a bond-charge attraction ( $W>0$) term
\begin{equation}
H_W=-W \sum_l \left( B_{l,l+1}\right) ^2 ~,
\end{equation}
with the bond-charge density operator $B_{l,l+1}$ defined as
\begin{equation}
B_{l,l+1} = \sum_{\sigma} \left( c_{l,\sigma}^{\dagger}c_{l+1,\sigma}
+ c_{l+1,\sigma}^{\dagger}c_{l,\sigma} \right).
\end{equation}
The bond-charge attraction $H_W$ can be obtained from a Su-Schrieffer-Heeger
e-p interaction \cite{18.Su} by integrating out the phonon degree of freedom
in the antiadiabatic limit ($M\rightarrow 0$ ) \cite{19.Fradkin}. So the
W-term could also be viewed as a contribution of the e-p interaction with the
parameter related to the dimensionless e-p coupling \cite{18.Su} by $W=\pi
\lambda t/4$ and the Hamiltonian (1) has included both the quantum phonon
fluctuation and the electron correlation for a one-dimensional
electron-phonon interacting system.

First, let us consider the symmetry properties of the model (1).
It is obvious that the model has a usual SU(2) rotational symmetry in spin
space. The generators of the rotations are

\begin{equation}
S^z = \frac 12 \sum_i \left( c_{i,\uparrow }^{+}c_{i,\uparrow}
-c_{i,\downarrow }^{+}c_{i,\downarrow }\right) ,
S^{+} = \sum_i c_{i,\uparrow }^{+}c_{i,\downarrow } ,
S^{-} = \sum_i c_{i,\downarrow }^{+}c_{i,\uparrow } ,
\end{equation}
which commute with the Hamiltonian (1). There is also an SU(2) axial charge
symmetry as in the Kondo model \cite{20.Tsunetsugu} and the Hubbard model 
\cite{21.Zhang}. The axial charge generators are

\begin{equation}
I^z = \frac 12 \sum_i \left( c_{i,\uparrow }^{\dagger}c_{i,\uparrow }
+c_{i,\downarrow }^{\dagger}c_{i,\downarrow }-1\right) ,
I^{+} = \sum_i \left(-1\right)^i c_{i,\uparrow }^{\dagger}c_{i,\downarrow } ,
I^{-} = \sum_i \left(-1\right)^i c_{i,\downarrow }^{\dagger}c_{i,\uparrow } ,
\end{equation}
which have the same commutation relation as the spin operators in Eq.(5). So
the Hamiltonian (1) has an SU(2)$\times $SU(2) symmetry group as the
alternating Hubbard model \cite{22.Pang}. It is clear that the $z$ component
of the axial charge operator is nothing but the total electron
number operator, while its transverse components are the staggered pairing
operators, called $\eta $-pairing. \cite{23.Yang}

Due to the commutation between the spin and charge generators, we could
identify the eigenstates of the Hamiltonian (1) by the eigenvalues of
the total spin $S^z$ and total charge operators $I^z$ $\mid S^z,I^z\rangle$,
then the spin gap and charge gap are defined as

\begin{equation}
\Delta _s=E_0\left( 1,0\right) -E_0\left( 0,0\right) ,\Delta _c=E_0\left(
0,1\right) -E_0\left( 0,0\right)
\end{equation}
where $E_0\left( S^z,I^z\right) $ is the lowest energy in 
the $\left(S^z,I^z\right) $ subspace, and it must have the eigenvalues for 
the spin $S=S^z$ and charge $I=I^z$ \cite{24.Xiang}.

The charge-spin transformation

\begin{equation}
c_{i,\uparrow }\rightarrow c_{i,\uparrow },c_{i,\downarrow }\rightarrow
\left( -1\right) ^ic_{i,\downarrow }^{\dagger}
\end{equation}
interchanges the spin generator (5) and charge generator (6),
hence the spin and charge gaps in Eq.(7).
It is straightforward to show that the bond-charge operator (4) is invariant
under the transformation that Hubbard repulsion turns into attraction .
Thus we have an equality for the spin and charge gaps

\begin{equation}
\Delta _s\left( t,W,U\right) =\Delta _c\left( t,W,-U\right)
\end{equation}
So if the spin gap is closed at a critical $U\left( >0\right) $, the charge
gap must be closed at $-U$, the system will be in a metallic phase.

Now, we come to the numerical treatment of the Hamiltonian (1) by the
powerful DMRG method \cite{14.White}, which allows an essentially exact
treatment of electron correlations at a fully quantum mechanical level. The
open boundary condition is adopted since it is usually performed much better
than the periodic boundary condition. The hopping integral $t$ is set to 1 as
the unit of energies. Up to $m=512$ states are tested and finally $m=128$
states are kept in the DMRG calculation for the systems with $L=48, 64, 80$, and
128, the truncation error is less than $10^{-9}-10^{-6}$ at all steps. After
four iterations, the change in the ground-state energy is at seven digits
between the successive iterations. We checked our DMRG calculations against
exact numerical results for long (up to 128 site) non-interacting ( $U=W=0$)
chains, excellent agreements were found both in the correlation
functions and in the energies of the ground and low-lying excited states.
The corresponding relation (9) between the spin and charge gaps is also
confirmed in our calculation within numerical error.
Moreover, we have carefully compared our charge gap with the Bethe Ansatz
exact solution of the Hubbard model \cite{25.Lieb} and obtained
quantitative agreement.

The first quantity we calculate is the spin gap, Eq. (7).
In the case of $U=0$, the bond-charge attraction $W$-term will induce a BOW
long-range order as the e-p interaction in the SSH model \cite{18.Su}. The
charge and spin gaps are equal in its excitation spectra. All fluctuations
besides the BOW are completely suppressed. In the presence of Hubbard
repulsion $U>0$, the charge gap increases while the spin gap decreases with
increasing $U$. Figure 1 shows the spin gap $\Delta _s$ versus the Hubbard $U
$ for different values of the bond-charge attraction $W$. The inset shows a
linear extrapolate of $\Delta _s\left( L\right) $ with the inverse of chain
length $1/\left( L+1\right) $. The spin gap $\Delta _s$ at thermodynamic
limit in Fig. 1 decreases monotonically with the increase of $U$.

Within a g-ology continuum-limit model, Emery \cite{26.Emery} has given out the
phase diagram for an interacting fermion system in the weak coupling region by
using the renormalization group method. The interaction parameters in the
Hamiltonian (1) can be transformed into the g-ology ones,
\begin{equation}
g_1 = U - 8W, \hspace{1.0cm} g_2 = U, \hspace{1.0cm} g_3 = U + 8W.
\end{equation}
The ground state depends only on the sign of $g_1$. $g_1=0$ is a phase
transition point from a charge density wave (CDW) to a spin density wave (SDW),
where the BOW character was not considered.
Combining Emery's theory and bosonization method, Japaridze considered the BOW
order and confirmed the phase transition from a spin-gapped state to a gapless
spin fluid at $U=8W$. Our calculation on the spin gap
is in good agreement with that of the weak-coupling theory
\cite {26.Emery,15.Japaridze}. The exact transition point is not
easy to be identified numerically since the spin gap decreases
exponentially, which is the same as that in the spin-Peierls model.
\cite{5.Sandvik,8.Campbell}

We now investigate the nature of the BOW ordering by examining the
staggered bond charge correlation function

\begin{equation}
C_{BOW}\left( r\right) =\frac 1L\left( -1\right)^r 
\sum_i \left\langle B_{i,i+1}B_{i+r,i+r+1}\right\rangle
\end{equation}
The spin density wave (SDW) order can be described by a similar staggered
spin-spin correlation function

\begin{equation}
C_{SDW}\left( r\right) = \frac 1L\left( -1\right)^r 
\sum_i \left\langle s_i^zs_{i+r}^z\right\rangle
\end{equation}
where the z-component spin is defined as

\begin{equation}
s_i^z = {1 \over 2} ( n_{i,\uparrow }-n_{i,\downarrow } ) ~.
\end{equation}
As is well known, for a uniformed system, the bond charge $\left\langle
B_{i,i+1}\right\rangle $ is uniformly distributed along the chain, and a rapid
decrease in the bond charge correlation function (11) is expected, that is,
the BOW order is short-ranged. However, for a dimerized system, the bond
charge $\left\langle B_{i,i+1}\right\rangle $ shows an oscillation on bonds,
the staggered bond charge correlation function $C_{BOW}\left( r\right) $
will approach to a nonzero constant $\Delta _{BOW}$, which shows the
existence of a BOW long-range order. Usually, it should take a much longer
chain for the bond charge calculation to identify the BOW long-range order 
\cite{13.Qin}, however, the bond charge correlation will be saturated for a
chain with a hundred sites.

In Figure 2 we show both the staggered bond charge (a) and spin-spin (b)
correlation functions. From Fig. 2(a) we can see that the staggered BOW
correlation functions show a similar behavior for both the spin-gapped state
($U<8W$ ) and the gapless spin fluid state ($U>8W$ ). And at the same time,
from Fig. 2(b), we can see that the spin-spin correlations show a different
behavior in these two regimes, it has a short-range SDW order in the
spin-gapped state ( $U<8W$) and a quasi-long-range SDW order in the
gapless spin fluid state ($U>8W$). The line in Fig. 2(b) is a 1/r fitting.
This result is in conflict with what Japaridze obtained in Ref. 15,
where it was pointed out that the phase transition at $U=8W$ is from
a spin-gapped state with a true long-range BOW order to
a gapless spin-fluid state characterized by an identical $1/r$ decay for both
quasi-long-range SDW and BOW orders. But our calculation indicates that the
BOW correlation shows a long-range order even when the spin gap is closed
in Fig. 2(a). While the obtained spin-spin correlation function can be
fitted very well by a 1/r-decay function (the line in Fig. 2(b)), the bond
charge correlation function in Fig. 2(a) obviously does not show the
1/r-decay behavior. It indicates that there is a novel dimerized spin fluid
phase for the Hamiltonian (1) at $U>8W$. By the charge-spin transformation
(8), the ground state will be in a dimerized metallic phase for $-U>8W>0$,
where the staggered BOW correlation shows a long-range order with a
vanished charge gap.

Figure 3 shows the BOW order parameter $\Delta _{BOW}$ versus the Hubbard
repulsion $U$ for different bond charge attraction $W$ for a $L=128$ chain.
In the weak bond charge interaction regime, the BOW order is first enhanced
by the Hubbard repulsion $U$ and reach a maximum,
then the BOW order is reduced, such behavior is similar to the e-p model
at adiabatic limit. \cite{3.Baeriswyl,4.Wu}
However, for a stronger bond charge interaction
(such as $W=0.2$), the BOW will be reduced monotonously by the Coulomb
interaction. When the Hubbard U is increased further, we will see that the
BOW order becomes almost independent of the bond charge interaction since the
Hubbard U dominates. There another phase transition is expected for the
BOW long-range order since the spin-Peierls model shows the transition
\cite{5.Sandvik,6.Bursill,7.Uhrig} when the quantum fluctuations
are taken into account.

In summary, we have found a novel dimerized spin fluid phase for the 1d
Hubbard model with a bond charge attraction $W$-term. 
The result indicates that there are two phase transition points,
the first one is around $U=8W$ for the spin gap and 
the second one is at $U<\infty $ for the BOW long-range order. The
BOW long-range order is shown to exist even when the spin gap is closed when
$U>8W$. The charge-spin transformation gives that the ground state should be
in a dimerized metallic phase for $U>-8W$.

We would like to acknowledge the useful discussions with
Prof. X. Sun, Y.P. Wang, Z.Y. Weng, T. Xiang, and G.M. Zhang.
This work is supported by National Natural Science Foundation of China
(No. 19725414) and 
partially supported by a grant from the Research Grants Council of the HKSAR,
China (Project No. CUHK 4288/00P).

\begin{figure}
\caption{The spin gap $\Delta _s$ versus the Hubbard repulsion $U$ for
different values of the bond-charge attraction $W$. The inset shows a linear
extrapolation of $\Delta _s\left( L\right) $ with the inverse of the chain
length $1/\left( L+1\right) $. }
\end{figure}

\begin{figure}
\caption{(a) The staggered bond charge correlation function $C_{BOW}(r)$,
and (b) the staggered spin-spin correlation function $C_{SDW}(r)$
for the chain of length $L=128$.
The dashed line in (b) is a $1/r$ fit. $W/t = 0.1$. }
\end{figure}

\begin{figure}
\caption{The bond charge order parameter $\Delta _{BOW}$
versus the Hubbard repulsion $U$ for the chain of length $L=128$.}
\end{figure}
%

\end{document}